\documentclass[aps,pra,preprint]{revtex4}
\usepackage{amsmath}
\usepackage{epsfig,amssymb,amsfonts}
\usepackage{graphicx}
\usepackage{natbib}
\usepackage{hyperref}

\begin{document}

\title{Why Bohmian velocity might not be the only quantum velocity and the role of diffusion flux in quantum theory}

\author{Charalampos Antonakos}

\today

\begin{abstract}
In this short-length paper, we will present some math that play a central role in quantum hydrodynamics and were presented by Mita in the past. In this formulation of QM, a quantity is involved which is the diffusion flux. As we will see, this quantity plays a crucial role on the evolution of wave packets. More specifically, we will initially briefly explain some results derived in Mita's papers, by analyzing Gaussian and soliton wave packets, while we also focus on the role of diffusion flux in the evolution of the square wave packet. Also, a very brief discussion will be made about the possible presence of the osmotic velocity field in quantum dynamics.

\end{abstract}
\maketitle

\section{Introduction}
In \cite{mita2003dispersive} Mita had introduced the quantum momentum density which contains a probability flux $\overrightarrow{J}$ term and a diffusion term $\overrightarrow{D}$
$$\Psi^*(r,t)\hat{P}\Psi(r,t)=M\left(\overrightarrow{J}(r,t)+i\overrightarrow{D}(r,t)\right)$$
where $M$ is the mass of the quantum particle. Now, using $\hat{P}=-i\hbar \overrightarrow{\nabla}$, where $\hbar$ is the Planck's constant, we obtain:
\begin{equation} \label{eq:1}
    \overrightarrow{J}(r,t)=\frac{\hbar}{M}Im\left(\Psi^*(r,t)\overrightarrow{\nabla}\Psi(r,t)\right)
    \end{equation}
    and
    $$\overrightarrow{D}(r,t)=-\frac{\hbar}{M}Re\left(\Psi^*(r,t)\overrightarrow{\nabla}\Psi(r,t)\right)$$
    Using
$\Psi(r,t)=R(r,t)e^{iS(r,t)}$, we are able to derive:
\begin{equation} \label{eq:2}
\overrightarrow{D}(r,t)=-\frac{\hbar}{2M}\overrightarrow{\nabla}\rho(r,t)
\end{equation}
which is also the expression for the diffusion flux in classical diffusion. Mita also showed that both the probability and diffusion flux contribute to the total kinetic energy of the particle:
$$E_K=\frac{1}{2}M\left(\int_{V}{\frac{J^2(r,t)}{\rho(r,t)}}d\tau+\int_{V}{\frac{D^2(r,t)}{\rho(r,t)}}d \tau\right)$$ 

In contrast to Bohm's theory \cite{bohm1952suggested}, where the particle doesn't move in stationary states (suppose a 1D system) since $\overrightarrow{v}^{\Psi}=\frac{\overrightarrow{J}^{\Psi}}{\rho}=0$, here we have quantum fluctuations that lead to a quantum Brownian motion. In other words, there is still contribution to the total kinetic energy due to diffusion. This kinetic energy is associated with the momentum uncertainty of the particle due to fluctuations. More specifically:
\begin{equation} \label{eq:3}
E_{K,x}^{(stationary)}=\frac{(\Delta P)^2}{2M}=\frac{1}{2}M\int_{\mathbb{R}}{\frac{D^2(x)}{\rho(x)}}dx
\end{equation}
\textit{The diffusion flux, in general, is an indicator of the tendency of the wavepacket to diffuse.}

\section{THE OSMOTIC VELOCITY}

We should note that, in the past, there was also a quantity named osmotic velocity that was discussed by Bohm and Hiley \cite{bohm1989non} in the context of a stochastic version of Bohmian mechanics. This velocity was defined as follows: $$\overrightarrow{u}(r,t)=\frac{\hbar}{2M}\frac{\overrightarrow{\nabla}\rho(r,t)}{\rho(r,t)}=-\frac{\overrightarrow{D}(r,t)}{\rho(r,t)}$$. 
The role of that velocity according to the authors was to balance diffusion in stationary state systems, where there was no probability evolution over time. This velocity, according to the authors, is responsible for "guiding" the particle in higher probability areas. It was also noted that the osmotic velocity becomes infinite at the nodes of the wave function. Based on that, we'll give our own interpretation of this quantity. To be more specific, we know that for every position square-integrable wave function $\Psi(x)$, we have a corresponding momentum wave function $\Psi(\rho)$ which is also square-integrable. This means that $|\Psi(\rho \rightarrow \infty)|^2=0$. In other words, it's very difficult to find a particle that has extremely large velocity. And indeed, this is what we observe with the osmotic velocity. \textit{The particle has very large osmotic velocity close to the nodes of the wave function, where it is obviously very difficult to be observed. And that, of course, sounds very natural since the dwell time of a particle at a specific spatial region is inversely proportional to its velocity in that region} (there is no boundary in the velocity since we are talking about non-relativistic QM). In a similar way, we can explain why the osmotic velocity approaches zero in the highest probability areas. And, of course, by taking into advantage the osmotic velocity expression, Eq. (3) gives us in stationary states that $E_{K,x}^{stationary}=\frac{1}{2}M\int_{\mathbb{R}}\rho(x)u^2(x)dx=\frac{1}{2}M<u^2(x)>$, as expected.

Now that we are finished with the osmotic velocity, we will move on to the analysis of the evolution of wave packets. In chapter III we will briefly discuss both Gaussian and soliton wave packets and at the end, in Chapter IV, we will see the role of diffusion flux in the evolution of initially extremely localized quantum systems.

\section{DIFFUSION FLUX IN GAUSSIAN WAVE PACKETS AND SOLITONS}

The role of this Chapter is to introduce the diffusion flux by looking specifically at the evolution of two types of waveforms. First, we have the Gaussian wave packet:
$$\Psi_g(x,t)=\left(\frac{2}{\pi}\right)^{\frac{1}{4}}\frac{1}{\sqrt{\pi}}e^{-(x-u_{0}t)^2/\epsilon^2}e^{i(k_0x-\frac{\hbar k_0^2}{2m}t+\delta)}$$  where $\epsilon(t)=\alpha\sqrt{1+\left(\frac{t}{T}\right)^2}$ , $u_0=\frac{\hbar k_0}{m}$ , $T=\frac{ma^2}{2\hbar}$  and  
 $$\delta=\frac{(x-u_0t)^2}{\epsilon^2}\frac{t}{T}-\frac{1}{2}tan^{-1}\left(\frac{t}{T}\right)$$

If we use Eqs. (1) and (2) according to Mita \cite{mita2021schrodinger} we get
$$\overrightarrow{J}_g(x,t)=\rho_g(x,t)u_0\overrightarrow{e}_x+\frac{t}{T}\overrightarrow{D}_g(x,t)$$ 

The first term is related to the "centre of mass" motion of the wave packet and the second term, which contains the diffusion flux is related to the expansion of the probability distribution. After very long time, the system approaches an equilibrium state in which we observe that the wave packet diffusive tendency approaches zero. Indeed, let's choose for simplicity that $u_0=0$ and assume that $t>>0$ and $|x|<<\infty$. We observe that since $$\overrightarrow{D}_g(x,t)=\frac{2\sqrt{2}\hbar}{\sqrt{\pi}m\epsilon^3(t)}xe^{-2x^2/\epsilon^2(t)}\overrightarrow{e}_x$$ we obtain:
$$\overrightarrow{D}_g(x,t>>0) \approx\left(\frac{2 \sqrt{2}\hbar}{\sqrt{\pi}ma^3}\frac{xT^3}{t^3}\left(1-2\frac{x^2T^2}{a^2t^2}+O\left(\frac{x^4}{t^4}\right)\right)\right)\overrightarrow{e}_x\Rightarrow$$
$$\overrightarrow{D}_g(x,t>>0) \approx\left(\frac{2 \sqrt{2}\hbar}{\sqrt{\pi}ma^3}\frac{xT^3}{t^3}+O\left(\frac{x^3}{t^5}\right)\right)\overrightarrow{e}_x\Rightarrow$$
$$\overrightarrow{D}_g(x,t>>0)\sim \frac{x}{t^3}\overrightarrow{e}_x=s\overrightarrow{e}_x$$ where $|s|<<1$. Also, $|\overrightarrow{D}_g(x,t=t_2>>0)|<|\overrightarrow{D}_g(x,t=t_1>>0)|$, where $t_1<t_2$ implying that the diffusion tendency decreases with time for large enough times.
Now, when it comes to the soliton wave packet, Mita \cite{mita2021schrodinger} defined a probability distribution:
$$\rho_{sol}(x,t)=\frac{1}{2\sigma_0}sech^2\left(\frac{x-u_0t}{\sigma_0}\right)$$
and found the potential $V(x,t)$ so that $\rho_{sol}$ behaves as a soliton. He found that:
\begin{equation} \label{eq:4}
V(x,t)=-\frac{2\hbar^2}{m \sigma_0}\rho_{sol}(x,t)
\end{equation}
At first sight, from a hydrodynamical perspective, this equation doesn't give us any physical intuition. However, what we will do is to write the equation in the following form:
\begin{equation} \label{eq:5}
\overrightarrow{F}(x,t)=\frac{2\hbar^2}{m\sigma_0}\overrightarrow{\nabla}\rho_{sol}(x,t)\propto-\overrightarrow{D}_{sol}(x,t)
\end{equation}
And that's how the form of the wave packet is preserved. The greater the tendency of the wave packet to diffuse (greater $\overrightarrow{D}$) in a specific direction, the larger the "effective" force of opposite direction that has to be applied. And, of course, the coefficient $\mu=\frac{2\hbar^2}{m\sigma_0}$ is position-independent, because otherwise it would cause deformation of the wave packet.

We have to add that this scalar "effective" potential we referred to is present in the very well-known NLS equation for solitons \cite{zakharov1974complete}, which appears a lot in Bose-Einstein condensates:
$$i\hbar \partial_t\Psi_{sol}(x,t)=-\frac{\hbar^2}{2m}\partial_x^2\Psi_{sol}(x,t)-\mu |\Psi_{sol}(x,t)|^2\Psi_{sol}(x,t)$$

It must be remarked that the Eq. (5) appears to be the fundamental condition for production of soliton solutions rather than Eq. (4) since the transformation $V'(x,t)=V(x,t)+\gamma(t)$, under which Eq. (5) remains invariant, produces solutions of the form:
$$\Psi'_{sol}(x,t)=\Psi_{sol}(x,t)e^{-i\gamma (t)/\hbar }$$
that behave like solitons too.
\section{DIFFUSION AND SQUARE WAVE PACKETS}

In this Chapter, we will examine a state that at $t=0$ is extremely localized. More specifically, we will consider the initial state: $$\Psi(x,0)=\frac{1}{\sqrt{a}}\chi_{[-a/2,a/2]}(x)$$
According to Mita again \cite{mita2007dispersion}, the time-evolved state in free space will be the following:
$$\Psi(x,t)=\frac{(-1)^{3/4}}{2\sqrt{ia}}\left[ erfi \left[(-1)^{1/4}\sqrt{\frac{m}{2 \pi \hbar t}}(x-a/2)\right]-erfi\left[(-1)^{1/4}\sqrt{\frac{m}{2 \pi \hbar t}}(x+a/2)\right]\right]$$
After very short time and very far very away from $x=a/2$, we have that:
$$\rho(x>>a/2,t<<m/2\pi \hbar)\sim \left|\frac{e^{z_1^2}}{\sqrt{\pi}z_1}\left(1+\frac{1}{2z_1^2}+O\left(\frac{1}{z_1^4}\right)\right)-\frac{e^{z_2^2}}{\sqrt{\pi}z_2}\left(1+\frac{1}{2z_2^2}+O\left(\frac{1}{z_2^4}\right)\right)\right|^2$$
where $z_1=\sqrt{\frac{m}{2i\hbar t}}(x-a/2)$ and $z_2=\sqrt{\frac{m}{2i\hbar t}}(x+a/2)$
From above, we can conclude that 
$$Prob(x\in [L_1,L_2] ,t<<m/2\pi\hbar|L_2,L_1>>a/2)\approx \int_{L_1}^{L_2}\rho(x>>a/2,t<<m/2\pi \hbar)dx \neq 0$$
This reveals  a non-zero probability of finding the particle infinitely far away from $x=a/2$ after very small time. But, hydrodynamically speaking, what is the cause of that infinite diffusion of the wave packet? The answer is the infinite diffusion flux for the initial state at the points $x=-a/2$ and $x=a/2$, that reveals the infinite diffusion tendency of the initial wave packet at those points. The infinite suppression of the initial wave packet causes enormous diffusive tendency and this manifests itself after the removal of the physical "walls" that restrict the wave packet. More specifically, we have:
$$\Psi(x,0)=\frac{1}{\sqrt{a}}\Theta(a/2-|x|)=\frac{1}{\sqrt{a}}\Theta(a/2-x)\Theta(a/2+x)$$

By applying Eq. (2) to the system and given that $\frac{d\Theta(x)}{dx}=\delta(x)$, as well as $\Theta^2(x)=\Theta(x)$ we obtain the expression:
$$\overrightarrow{D}(x,0)=-\frac{\hbar}{2ma}\left[\delta(a/2+x)\Theta(a/2-x)-\delta(a/2-x)\Theta(a/2+x)\right]\overrightarrow{e}_x$$
where $\overrightarrow{e}_x$ a unit vector that points to the positive $x-$direction.
We observe that
$$\lim_{x\rightarrow a/2^-}\overrightarrow{D}(x,0)=(+\infty)\overrightarrow{e}_x$$
and 
$$\lim_{x\rightarrow -a/2^+}\overrightarrow{D}(x,0)=(-\infty)\overrightarrow{e}_x$$
as expected.

\section{DISCUSSIONS AND CONCLUSIONS}

In this paper, we managed to give some clues to support the idea of contribution of the osmotic velocity field on quantum dynamics, even on stationary states where Bohm's/current velocity plays no role. We discussed the reasons behind the relaxation to equilibrium of Gaussian wave packets after very long time and explained solitons's behavior as a result of counterbalance between diffusion flux and effective forces. We also presented  the mathematical and physical/hydrodynamical reason behind the infinite diffusion of strictly localized square wave packets. So, one the one hand, if someone wants to examine the evolution of a wave-packet, they'd better look at the Bohmian current velocity, but on the other hand, as we saw, diffusion flux also provides us with significant information relevant to the origin of quantum diffusion, which arises from the diffusion tendencies of the initial state (that are expressed by the diffusion flux).

{}


\begin{thebibliography}{}

\bibitem{mita2003dispersive} Dispersive properties of probability densities in quantum mechanics, Mita, Katsunori, American Journal of Physics {\bf 71} 894--902 (2003)

\bibitem{bohm1952suggested} A suggested interpretation of the quantum theory in terms of" hidden" variables. I, Bohm, David, Physical review {\bf 85} 166 (1952)

\bibitem{bohm1989non} Non-locality and locality in the stochastic interpretation of quantum mechanics, Bohm, David and Hiley, Basil J, Physics Reports {\bf 172} 93--122 (1989)

\bibitem{mita2021schrodinger} Schr{\"o}dinger's equation as a diffusion equation, Mita, Katsunori, American Journal of Physics {\bf 89} 500--510 (2021)

\bibitem{zakharov1974complete} On the complete integrability of a nonlinear Schr{\"o}dinger equation, Zakharov, Vladimir Evgen’evich and Manakov, Sergei Valentinovich, Theoretical and Mathematical Physics {\bf 19} 551--559 (1974)

\bibitem{mita2007dispersion} Dispersion of non-Gaussian free particle wave packets, Mita, Katsunori, American Journal of Physics {\bf 75} 950--953 (2007)

\end{thebibliography}
\end{document}